\documentclass[10pt,letterpaper]{article}
\usepackage{opex3}

\begin{document}

\title{Selective excitation of bright and dark plasmonic resonances of single gold nanorods}

\author{O. Demichel, M. Petit, G. Colas des Francs, A. Bouhelier, E. Hertz, F. Billard, F. de Fornel, B. Cluzel}
\vspace{5pt}

\email{benoit.cluzel@u-bourgogne.fr}
\address{Laboratoire Interdisciplinaire Carnot de Bourgogne, CNRS-UMR 6303, Universit\'e de Bourgogne, 21078 Dijon, France}

\begin{abstract}
Plasmonic dark modes are pure near-field resonances since their dipole moments are vanishing in far field. These modes are particularly interesting to enhance nonlinear light-matter interaction at the nanometer scale because radiative losses are mitigated therefore increasing the intrinsic lifetime of the resonances. However, the excitation of dark modes by standard far field approaches is generally inefficient because the symmetry of the electromagnetic near-field distribution has a poor overlap with the excitation field. Here, we demonstrate the selective optical excitation of bright and dark plasmonic modes of single gold nanorods by spatial phase-shaping the excitation beam. Using two-photon luminescence measurements, we unambiguously identify the symmetry and the order of the emitting modes and analyze their angular distribution by Fourier-space imaging. \end{abstract}

\bibliographystyle{osajnl}

\section*{Introduction}
Localized surface plasmons (LSPs) are quantized collective oscillations of conduction electrons sustained by metal nanoparticles. The electromagnetic fields associated with LSPs are typically enhanced and strongly confined to the surface. Manipulation of the surface plasmons triggered a paradigm to control light-matter interactions at the nanoscale by a proper engineering of the surface~\cite{Aizpurua2005PRB,Schuck2005PRL,Miyazaki2006PRL,Stockman2004PRL}. The development of optical antennas~\cite{Novotny2007PRL} is largely relying on such engineering with benchmark applications ranging from surface-enhanced spectroscopies~\cite{Wang2006AdvMat_SERS} to local manipulation~\cite{Righini2009NL,Novotny1997PRL}, and molecular control~\cite{Gordon2012NL,Vanhulst1998BioImag,Busson2012NatComm,Gerard2008OptExpr}.

LSP responses  are generally inferred from the multipolar nature of the charge oscillations~\cite{Aizpurua2005PRB,Nordlander2004NL}. Depending on the  symmetry of the modes, LSPs may have a dipole-active character and are considered as bright modes because they efficiently decay in the far field. Maxwell's equations also predict the existence of dipolar-inactive LSPs : quadrupolar and higher-order eigenmodes cannot interact \textit{classically} with light and they are generally referred as dark modes~\cite{Nordlander2004NL,Liu2009PRL}.  An interesting consequence is that the intrinsic lifetime of the mode is typically longer than radiative modes, providing thus enhanced near-field interactions with single emitters~\cite{Liu2009PRL} or other plasmonic elements. As an example, Solis \textit{et al.} showed that light propagation through nanoparticle assemblies via dark LSPs is ten fold more efficient than with bright SPs~\cite{Solis2012NL}. 

Dark resonances may arise when nanoparticles are paired. Near-field coupling results to an hybridization of the system with the appearance of  bright bonding modes and a dark antibonding modes in the gap. The spatial distribution of these dark modes were mapped by detecting nonlinear photoluminescence with focused fields~\cite{Huang2010NL}, phase-shaped excitations~\cite{Volpe2009NL,Gomez2013NL} as well as second-harmonic generation~\cite{Berthelot2012OptExpr}. At the level of an isolated plasmonic structure, an optical excitation of dark LSPs is symmetry-forbidden for a normally incident plane-wave excitation.  The spectral signature of dark modes was however reported with an oblique incidence~\cite{Capasso2006APL,Ditlbacher2005PRL} and their spatial distribution interrogated by electron-energy loss spectroscopy and cathodoluminescence experiments~\cite{Chu2009NL,Schmidt2012NL}. 

In this letter, we explore the bright and dark modal distributions sustained by isolated single gold nanorods as a function of their length. A two-photon luminescence  non linear process (TPL) is used to probe the local electromagnetic field enhancement provided by LSP resonances~\cite{Bouhelier2003APL,Chicanne2002PRL,Muhlschlegel2005Sc,Imura2005JPhysChmeB}. Specifically, we demonstrate a selective excitation of bright and dark SP resonances using a first-order Hermit-Gaussian beam. Experimental data are compared to numerically calculated near-field intensities and phase distributions. Finally, we demonstrate the dipolar and quadrupolar character of resonant bright and dark LSP modes by measuring the angular distribution of the back-scattered light at the excitation wavelength.

\section*{Experimental}
\begin{figure}[t]
\includegraphics*[width=0.8\textwidth]{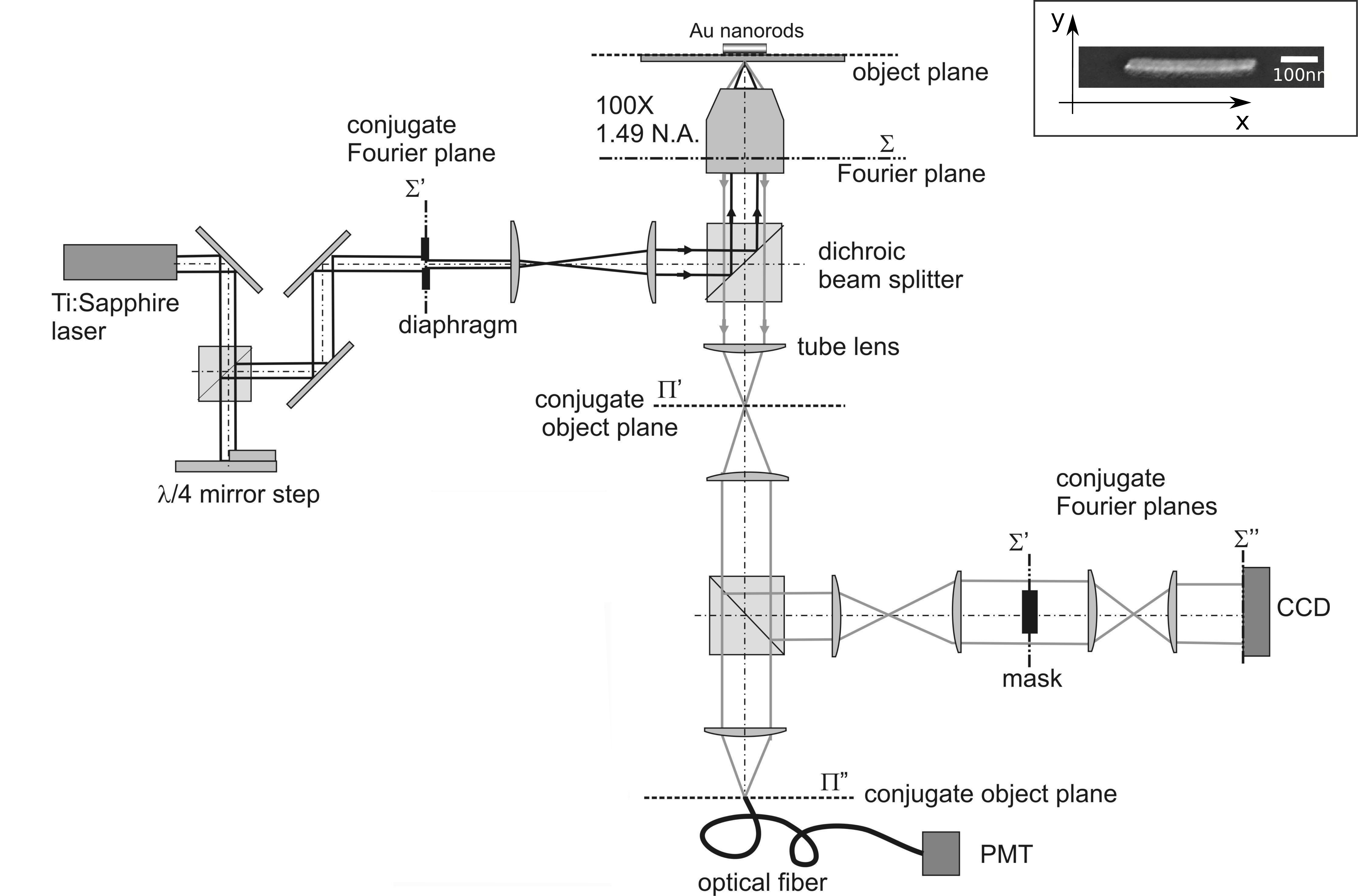}
\centering\caption{\label{setup_fig}  Schematic view of the experimental setup described in the text. In inset is shown a SEM image of a 500 nm long gold nanorod.}
\end{figure}

Nanorods were realized by electron beam lithography and gold evaporation on a conductive glass cover slip. The dimensions of the nanorods were 35~nm thick, 55~nm wide with length $L$ varying from 90~nm to 925~nm by steps of 15~nm. A scanning electron micrograph (SEM) of a 500 nm long nanorod is displayed in inset of  Fig.~\ref{setup_fig}.

TPL confocal images were recorded using the experimental arrangement sketched in Fig.~ \ref{setup_fig}. The patterned cover slip was placed on a piezo-electric scanning stage fitted on an inverted microscope. A Ti:Sapphire laser producing 120~fs pulses at a wavelength of 800~nm was focused on a diffraction-limited spot by high aperture microscope objective ($\times$60, numerical aperture NA=1.49). Two excitation configurations were used to selectively excite bright and dark LSPs. Bright modes were excited by a classical $HG_{0,0}(x,y)$ gaussian beam profile (axis $x$ and $y$ are defined in inset of figure \ref{setup_fig}). Dark modes were generated by a $HG_{1,0}(x,y)$ beam profile formed by centering a $\lambda$/4-step  gold mirror in the laser beam prior to the microscope objective~\cite{Novotny98UM}. The indices of these Hermitt-Gaussian beams correspond to the number of $\pi$-phase changes along the corresponding Cartesian direction. For the remaining of the discussion, the excitation polarization direction is fixed along the long nanorod axis ($x$ direction according to the inset of Fig. \ref{setup_fig}). The TPL signal generated when the nanorods are raster scanned through the waist of the focused beam was recorded by an amplified photo-multiplier tube (PMT) via a 200~$\mu$m optical fiber core. The TPL signal was filtered out with the help of a dichroic beamsplitter. To analyze the angular distribution at the excitation wavelength, a conjugate Fourier plane  of the microscope $\Sigma''$ was projected on a charge-coupled device (CCD) camera.  For these measurements, we reduced the excitation numerical aperture (NA$_{exc}<$1.0) by a diaphragm placed in a conjugate Fourier plane $\Sigma'$  before the microscope. In the detection path, a solid mask positioned in an intermediate Fourier plane $\Sigma'$ rejects the direct reflected light within the NA$_{exc}$ to only detect the scattering contribution distributed between the $NA_{exc}$ and the maximum NA, here 1.49~\cite{Huang2008PRB}. For Fourier analysis, a 50/50 mirror was used instead of the dichroic beamsplitter.

Numerical simulations were realized with a commercial Finite-Element-Method software. The full three-dimensional model was composed of an orthorhombic gold nanorod placed on a glass substrate with a dielectric constant $\epsilon_{g}$=2.25. The gold dielectric constants were taken from~\cite{johnson1972PRB}. Perfectly matched layers were added to the model to avoid any boundary reflections. The model was solved for an incident plane wave coming from the substrate with a propagating direction orthogonal to the glass surface. To take into account the fast decrease of the electromagnetic field around the nanorod, the discretization was finer than 2~nm.

\section*{Results and Discussion}
\begin{figure}[h!]
\includegraphics*[width=0.7\textwidth]{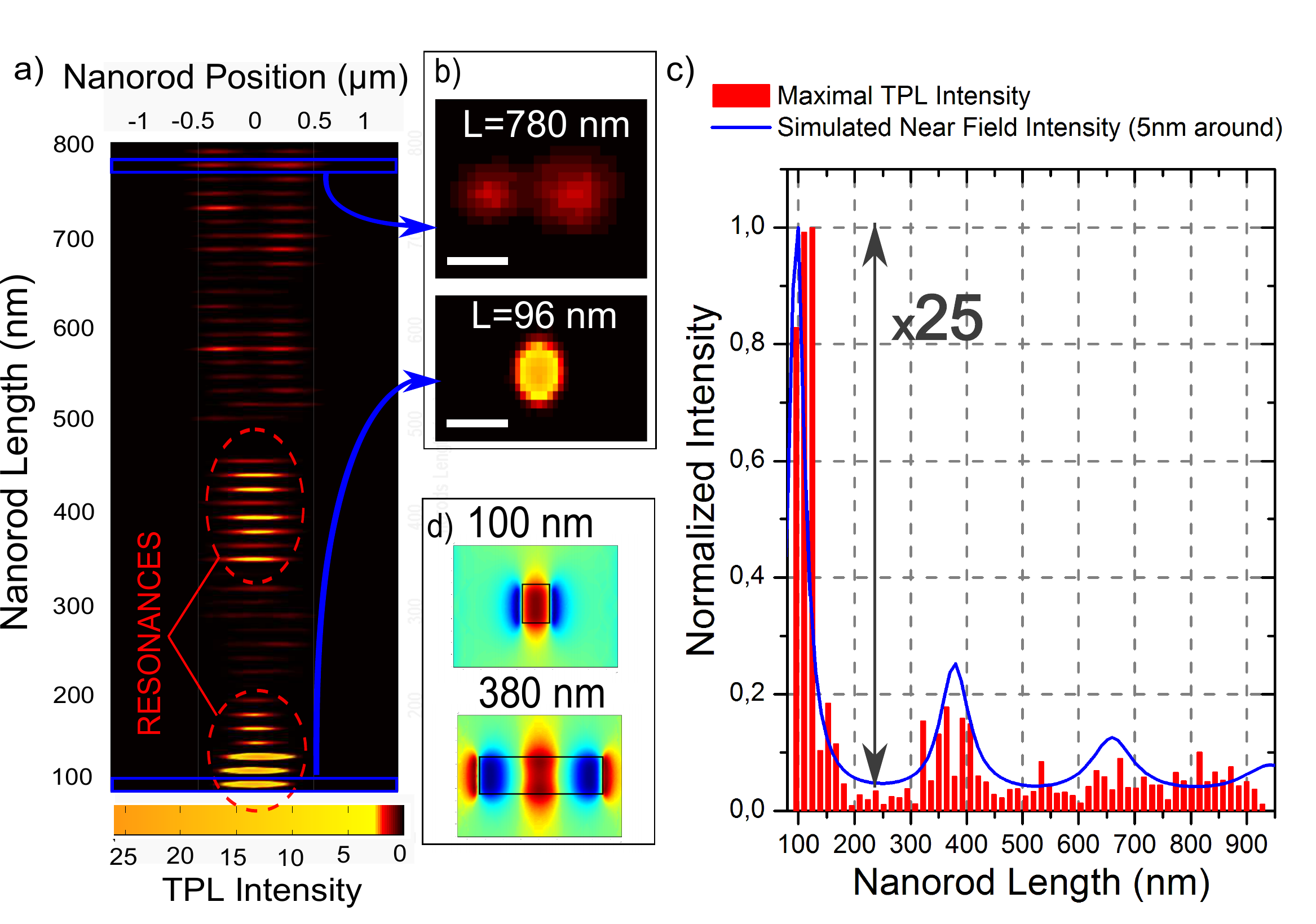}
\centering\caption{\label{HG00-fig} \textbf{Bright Modes :} (a) Distribution of TPL intensity versus nanorod length for a $HG_{0,0}(x,y)$ excitation beam. The horizontal axis corresponds to the nanorod position relatively to the excitation beam. (b) Zoom on the confocal images of the TPL from the 96 and 780 nm long nanorods. The scale bar is of 500 nm. (c) The length dependence of the maximum TPL signal (red columns) is compared to the simulated intensity of the electric field integrated 5 nm around the nanorods (blue line) for a plane wave excitation. (d) Simulated spatial distribution of the $E_x$ component 10~nm below the nanorod for a 100~nm and a 380~nm long nanorods at an excitation wavelength of 800~nm. The black lines indicate the contour of the corresponding nanorods. }
\end{figure}

Figure~\ref{HG00-fig}(a) shows  the experimental TPL intensity map as a function of nanorod length obtained for a $HG_{0,0}(x,y)$ excitation beam. The horizontal axis corresponds to the spatial position of the nanorods relatively to the focused beam center. Because TPL intensity has a quadratic dependence on the near-field intensity an increase signal relates to a local enhancement of the electric field; TPL measurements are thus an efficient tool to identify LSP resonances~\cite{Volpe2009NL}. The representation clearly shows an increase of the TPL intensity related to SP resonances for lengths comprised in the [95,130]~nm and [330,400]~nm ranges. Figure~\ref{HG00-fig}(b) reports two enlarged confocal maps for 96 and 780 nm long nanorods. Confocal maps result from the convolution of the spatial intensity distributions of the excitation beam and of the SP mode. These maps are far-field images and their resolution is then diffraction-limited to 270 nm. Thus, nanorods shorter than this latter length directly probe the excitation intensity profile. In view of the confocal map of the 96 nm long nanorod, the excitation appears to have a gaussian profile as expected for the related excitation conditions. For longer nanorods, the TPL is only observed at the nanorod extremities and results from a local enhancement of the electric field at the structure boundaries as shown for the 780 nm long nanorod.

In Fig.~\ref{HG00-fig}(a), only the length dependence of the TPL mapping is reported and Fig.~\ref{HG00-fig}(c) reports the length dependence of the maximum of TPL intensity. The first resonance occurs for length of 96 nm where the TPL signal is 25-fold higher than the one of off-resonance nanorods. The experimental data (red columns) are compared to simulations  (blue solid line). Calculations of the near-field of orthorhombic structures produce strong field enhancements at the corners due to the abruptness of the structure. In order to reduce the impact of this field enhancement compared to the exaltation due to surface plasmon resonances, we report here the mean electric field intensity integrated in a volume of 5 nm around the nanorods instead of its maximum. Simulations are realized with a plane wave illumination linearly polarized along the nanorods. Calculated plasmonic resonant lengths are in good agreement with those experimentally detected by TPL measurements. Then, we plot in Figure~\ref{HG00-fig}(d) the distribution of the simulated $E_x$ component  (oriented along the nanowire long axis) for the two resonant lengths: 100 and 380~nm. These maps show that, at resonances, $E_x$ is maximum at the center of the nanorod which ensures a good coupling with a focused $HG_{0,0}(x,y)$ laser beam once it is centered to the nanorod. Here, we precise that these near-field $E_x$ maps cannot be compared to confocal TPL maps. To be convinced, one just has to note that simulated near-field $E_x$ components oscillate on distances shorter than 130 nm whereas TPL maps have a diffraction-limited resolution of 270 nm since they are far-field images related to $E^4$. The resonance lengths (100 and 380 nm) correspond to the first ($m$=1) and third ($m$=3) order plasmonic modes since $E_x$ presents respectively one and three field maxima inside the nanorods (the m index is the ratio between the nanorod length and the effective wavelength as described below in the Fabry Perot description of resonances).  However, as a change of sign in the $E_x$ distribution corresponds to a $\pi$-phase change, these odd modes present two $\pi$-phase changes at their extremities, as previously reported ~\cite{Tatartschuk2009OE}. The \textit{Coupled Mode Theory} proposed by A. Yariv \cite{Yariv1973IEEE} says that optical modes are efficiently excited when the excitation intensity and phase spatial profiles match with those of the latter optical modes. Then, we believe that SP modes could be more efficiently excited with focused $HG_{2,0}(x,y)$ or  $HG_{4,0}(x,y)$ beams for which the phase-matching would be optimized. Here, resonances of longer nanorods with higher-order odd modes are not observed experimentally since the overlap with the $HG_{0,0}(x,y)$ is vanishing when the mode order ($m$) increases. 

\begin{figure}[h!]
\includegraphics*[width=0.8\textwidth]{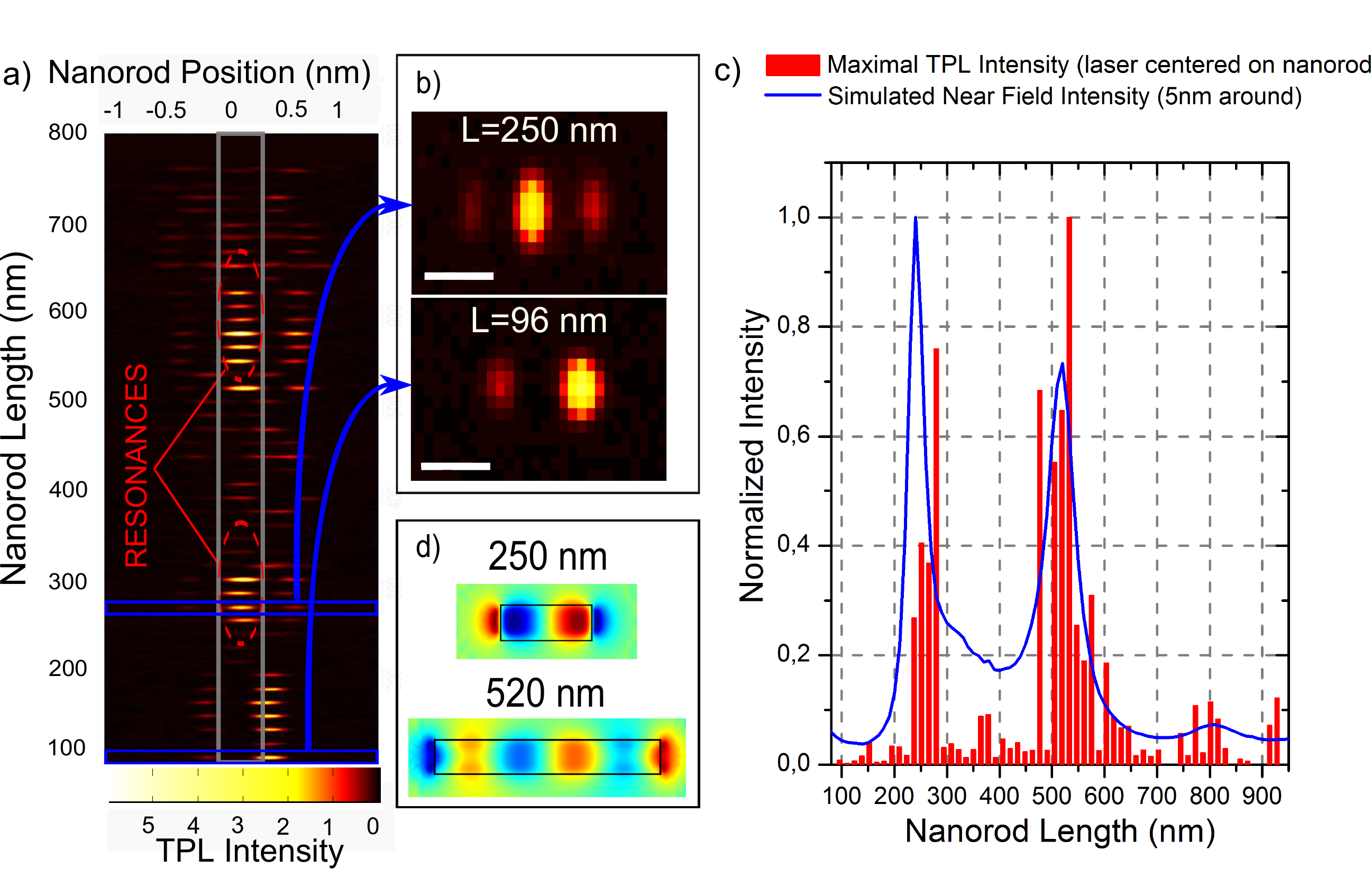}
\centering\caption{\label{HG01-fig}  \textbf{Dark Modes :} (a) Mapping of the nanorod TPL intensity versus the nanorod length for a $HG_{1,0}(x,y)$ excitation beam. The horizontal axis corresponds to the nanorod position relatively to the excitation beam. (b) Zoom on the confocal images of the TPL from the 96 and 250 nm long nanorods. The scale bar is of 500 nm. (c) The length dependence of the maximal TPL intensity (red columns) is compared to the simulated intensity of the electric field integrated in a volume of 5 nm around the nanorods (blue line). The excitation is similar to a $HG_{1,0}(x,y)$ focused beam centered on the nanorod. (d) Simulated spatial distribution of the $E_x$ component 10~nm below the nanorods for lengths of 250~nm and 520~nm at an excitation wavelength of 800~nm. }
\end{figure}

Next, we interrogate the TPL intensity for a $HG_{1,0}(x,y)$ excitation beam in order to excite even eigenmodes. Results are reported in Fig.~\ref{HG01-fig} which is organized in the same way than Fig.~\ref{HG00-fig}. Here, simulations are realized with an excitation similar to an $HG_{1,0}(x,y)$ focused beam. The TPL map of Fig.~\ref{HG01-fig}(a)  clearly shows a rise of the nonlinear response for a new set of length ranges: [240-300]~nm and [480-580]~nm. However, this figure is slightly more complex than Fig.~\ref{HG00-fig}(a). Indeed, as previously mentioned,  confocal maps result from the convolution of the spatial intensity distributions of the SP modes and of the excitation beam which is here modified by the spatial phase shaping. As above mentioned, shortest nanorods probe the spatial excitation intensity distribution. Fig.\ref{HG01-fig}(b) shows that the confocal TPL map of 96 nm long nanorod  is mainly composed of two lobes as expected for a focused $HG_{1,0}(x,y)$ beam. Note that the asymmetric lobe intensities result from imperfections of the home-made $\lambda$/4-step gold mirror.  Enlarged confocal map of the resonant 250 nm long nanorod exhibits three TPL maxima. The most intense one is obtained when the laser beam is centered on the nanorod, i.e. when the phase profiles of the excitation and of the plasmonic mode coincide. The two minor lateral spots are obtained when the nanorod extremities overlap each lobe of the focused $HG_{1,0}(x,y)$ beam. Even modes are thus efficiently excited only when the nanorod is centered with the focused $HG_{1,0}(x,y)$ laser beam.  For longer nanorods, confocal maps are becoming increasingly complex with four or even five lobes resulting from the convolution of confocal images similar to Fig.\ref{HG00-fig}(a) with the $HG_{10}(x,y)$ intensity distribution. The 
understanding of these complex maps is beyond the scope of this work. In the following, we consider the excitation of even modes occurring in the centered condition which is illustrated by the gray box of Fig.~\ref{HG01-fig}(a). Figure.~\ref{HG01-fig}(c) reports the maximal TPL intensity in that centered condition together with simulated near field intensities calculated with an excitation similar to a $HG_{1,0}(x,y)$ profile. Here again, there is a good agreement between the calculated length dependence of the electric field intensity (blue line) and the experimental data (red columns). Figure.~\ref{HG01-fig}(d) gives the simulated  $E_x$ spatial distribution for both resonant lengths: 250 and 520 nm. These modes are the second ($m$=2) and fourth ($m$=4) order eigenmodes since 2 and 4 maxima of $E_x$ are respectively present inside the nanorod. The dark nature of these modes is confirmed by the presence of a $\pi$-phase change at the center of the nanorods which also ensures a good match with the $HG_{1,0}(x,y)$ phase profile. Here, we unambiguously demonstrate the selective excitation of dark plasmonic modes by phase-shaping the excitation beam. One has to note that dark modes do not radiate by essence. Nevertheless, TPL which does not depend of the mode dipolar nature is still exalted at resonances (even if dark) since it is only sensitive to the local field intensity. TPL is then powerful to probe both bright and dark resonances of SPs. Finally, we anticipate that more complex phase-shaped beams (eventually produced by spatial light modulators) could be used to efficiently excite higher-order bright or dark modes.

LSP resonances can be described by the commonly accepted Fabry Perot resonator description of localized surface plasmon resonances~\cite{Novotny2007PRL,Ditlbacher2005PRL,Cubukcu2009APL}. Here, the fundamental propagating mode of a gold nanowire (not shown here) with a cross-section of 55~nm$\times$35~nm has a calculated effective index $n_{eff}$ of 3.03. The effective wavelength $\lambda_{eff}$ is thus $\lambda_0$/$n_{eff}$=264~nm. The tail of the field at the nanorod ends decays with a characteristic distance close to $\delta$=12~nm (resp 6, 4, 2.5 nm) for the $m$=1 (resp. $m$=2, 3, 4) mode (see Fig.\ref{HG00-fig}(d) and Fig.\ref{HG01-fig}(d)). In the above-mentioned description, the resonant modes are occurring when $L+2\delta=m\lambda_0/2n_{eff}$. Resonances are thus expected for lengths of $L$=108~nm, 252~nm, 388~nm, and 523~nm, in quite good agreement with the experimental four plasmonic resonances measured for lengths of $L$=100~nm, 250~nm, 380~nm and 520~nm.

\begin{figure}[t!]
\includegraphics*[width=0.95\textwidth]{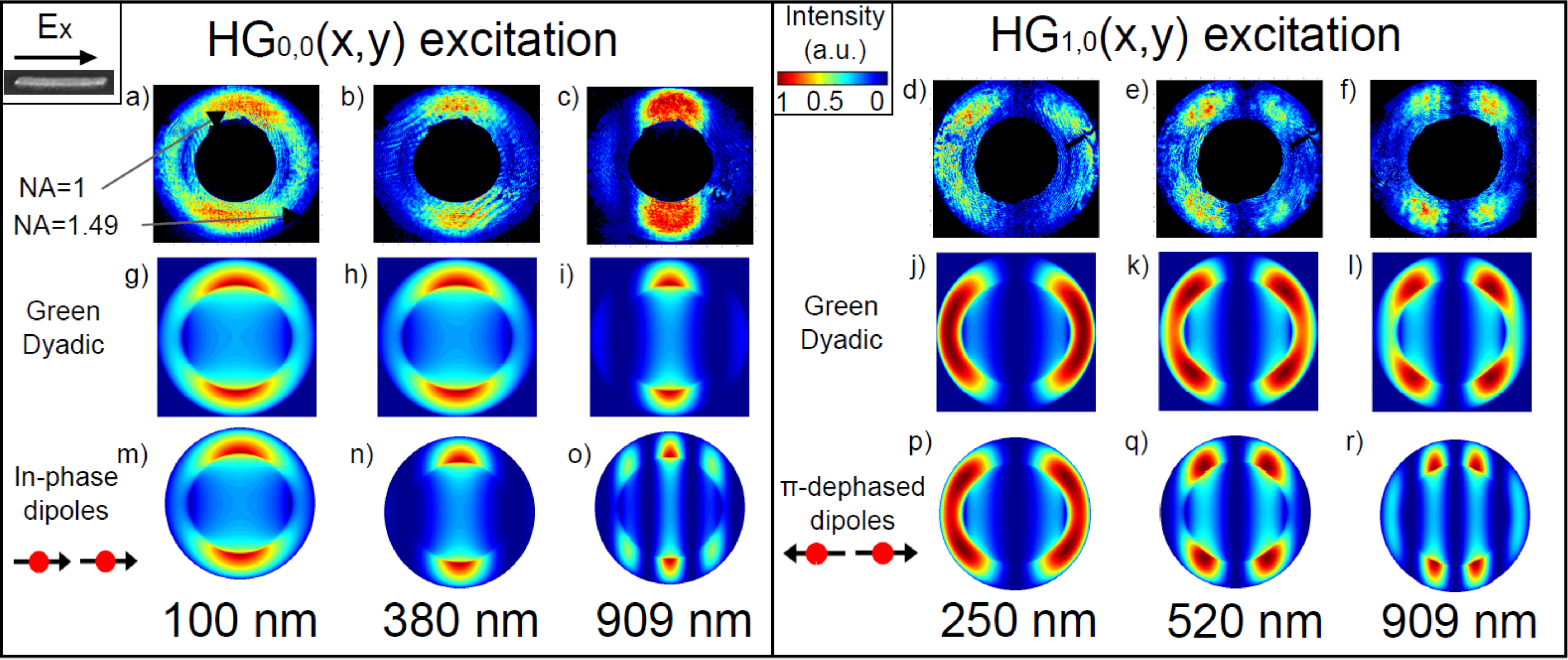}
\centering\caption{\label{Fourier-fig} (a-e) angular distributions (Fourier plane) of the back-scattered light polarized along the nanorod for $m$=1 to $m$=4 modes, and for an off-resonant 909 nm long structure. The detection is limited by the NA of the objective (1.49). The critical angle at NA=1 is also marked.  The filtering mask in clearly observed at the center of the Fourier plane images. (g-l) angular distributions calculated with Dyadic Green function for above mentioned cases. (m-r) angular distributions calculated by the coherent superposition of the field emitted by two in-phase or $\pi$-dephased dipoles separated by the corresponding nanorod lengths.}
\end{figure}

Now we turn to a far-field analysis of scattering properties of the modes by Fourier imaging. Bright and dark LSPs differ by their near-field distributions, and also by their far field scattering properties which are also dictated by the order of the modes~\cite{Taminiau2011NL,Curto2013NatComm}. In Fig.~\ref{Fourier-fig}(a-e) we report the experimental Fourier-plane images  of the back-scattered light for the four resonant modes above defined and for an out-of-resonance 909 nm long nanorod. Detection is polarized along the $E_x$ component (parallel to the nanorod long axis). In these images, the central black circle corresponds to the solid mask placed in the detection path way to reject the light distributed within the excitation numerical aperture. Figures~\ref{Fourier-fig} (a), (b) and (c) concern the $HG_{0,0}(x,y)$ excitation of the $m$=1 and $m$=3 bright LSP resonances and of the non resonant nanorod. These Fourier-plane images show that the bright modes principally radiate in a direction orthogonal to the nanorod long axis as expected from a dipolar emission. In particular, shortest nanorods have a dipolar-like pattern whereas the longest one exhibit fringes due to interferences originating from light scattered from different nanorod positions. It results from these interferences that the central fringe is narrower for longer nanorods and the directivity is then increased. 

Following the same approach, the images of Fig.~\ref{Fourier-fig} (d) , (e) and (f) concern the $HG_{1,0}(x,y)$ excitation of  the $m$=2 and $m$=4 dark LSP resonances and from the non resonant nanorod. It is worth to note that dark modes poorly scatter in far-field, and the experimental signal from these modes is at least ten times lower than bright modes. More precisely, for antenna shorter than 500 nm, we only collect signal from resonant antenna. The Fourier images of Fig.~\ref{Fourier-fig} (d) , (e) and (f) clearly differ from those of bright modes. In particular, dark modes present a minimum of emission exactly where bright modes principally emit. The orthogonality of the local field distribution between these modes appears here: in this plane orthogonal to the nanorod axis, bright modes constructively interfere whereas dark modes destructively interfere. Indeed, the $HG_{1,0}(x,y)$ excitation generates out-of-phase plasmon in both nanorod sides. Furthermore, it is especially interesting to note that the directivity of the emission from 250 nm long resonant nanorod is tilted by 90 degrees from the directivity of brights modes. Curto \textit{et al}. reported similar emission patterns resulting from a near-field excitation of single nanorods with a quantum dot placed in the vicinity of plasmonic elements \cite{Curto2013NatComm}. In the present work, the radiation patterns are obtained by scattering the laser field onto the antennas. Since the same modes are excited, the angular symmetry of the scattered radiation is the same as in \cite{Curto2013NatComm}. However, the origin of the signal is fundamentally different. With our approach, we demonstrate that the control of the spatial phase-profile of the optical excitation is a powerful tool to engineer the directivity of an optical antenna. Furthermore, from our measurements it seems that in the Fourier images do not conserve information on the mode orders and are only governed by the nanorod diffraction figures. More precisely, resonant nanorods have emission patterns similar to off-resonance nanorods of similar lengths (not shown) and only the intensity is modified. Here, the Fourier images are only sensitive to the parity of the excited mode.

Experimental emission patterns are also confronted to th ose calculated using the Green's dyad technique. To this aim, we first calculate the electric field inside the nanostructure as described in \cite{Yacoub2014PIER}. For simplicity, we use a plane wave excitation to simulate the $HG_{0,0}(x,y)$ excitation and a plane wave excitation but with a $\pi$-phase shift for negative x to compare to the  $HG_{1,0}(x,y)$ excitation. Then the intensity scattered in the Fourier plane is calculated using the asymptotic expansion of the Green's dyad \cite{Colas2002JPhysChem,Lieb2004JOptSAB}. Calculated Fourier-plane images are reported in Fig.~\ref{Fourier-fig} (g-l). The main features of experimental images: 1) the increase in directivity for longest nanorods, 2) the change in orientation of emission from bright to dark modes and 3) the zero emission in the nanorod perpendicular plane for dark modes are well described by the calculations. This technique also describes the fine structure of the emission patterns of the $HG_{0,0}(x,y)$ excited longest nanorods for which two lobes appear on both sides of the main fringe. If this technique is in good agreement with experimental measurements and allows to anticipate the emission properties of plasmonic nano-antenna, it is highly time consuming especially for multi-element and large scale antenna like Yagi-Uda. 

Then we propose a simple model to describe the nanorod emission pattern which is only based on the emission of two dipoles. In this basic model only two dipoles are considered. Depending on the dark or bright mode, they oscillate in phase or they are $\pi$-dephased as depicted at the left of figure~\ref{Fourier-fig}(m) and (p). Dipoles are placed 10 nm above a silica substrate (n=1.5). Figure~\ref{Fourier-fig}(m-o) (resp. (p-r)) report the calculated Fourier-plane images for in-phase (resp. out-of-phase) dipoles separated by the corresponding nanorod lengths. Qualitatively, this quite simple model reproduces very well the main features of the experimental patterns especially for shortest antenna. The accurate comparison with experimental emission patterns of bright and dark modes of short antenna is quite surprising and gives us a better understanding on the origin of the angular emission patterns. Indeed, the physical meaning of this two-dipole based model corresponds to a Fourier-plane pattern essentially governed by radiations from the nanorod extremities. It indicates that the radiative losses along the nanorods can be neglected for antenna shorter than 700 nm and only extremities emissions can be considered. Then, bright or dark modes only differ from the phase between extremities emissions. For longer antenna, the radiations all along the nanorod start to take an important role in the scattering patterns and have to be taken into account. The resulting patterns are then governed by diffraction figures as reported by Sersic $et\ al$ \cite{Sersic2011NJPhys}. However, we anticipate that the simplicity of this two-dipoles model can help to estimate the emission patterns of complex multi-element structures interacting through bright and/or dark modes. Furthermore, it has the advantage to propose analytical solutions and is poorly time-consuming compared to full-3D method like Green or FEM methods.

\section*{Conclusion}
To summarize, we used tightly focused Hermit-Gaussian modes  to selectively excite bright and dark plasmonic eigenmodes sustained by individual gold nanorods. Non-linear photoluminescence was used as probe of the local field maxima generated by the even or odd natures of the modes. In particular, we were able to unambiguously demonstrate the excitation of dark modes by introducing a $\pi$-shift in the excitation beam. We further showed that the emission pattern features the symmetry of the mode and can be qualitatively understood by considering two dipoles oscillating in phase or out of phase at the nanorod extremities. This work provides a simple avenue for selectively exciting bright or dark resonances that can be subsequently used to control the plasmonic response of single optical antennas.

\subsection*{Acknowledgments}
This project has been performed in cooperation with the Labex ACTION program (contract ANR-11-LABX-01-01). The research leading to these results has received funding from the R\'egion de Bourgogne under the PARI initiative, the PRES Bourgogne/Franche-Comt\'e  and the European Research Council under the European Community's Seventh Framework Program FP7/2007-2013 Grant Agreement no 306772. Calculations were performed
using DSI-CCUB resources (Universit\'e de Bourgogne).

\end{document}